\documentstyle[11pt]{article}
 
\begin{document}

\baselineskip=14pt plus 0.2pt minus 0.2pt
\lineskip=14pt plus 0.2pt minus 0.2pt


\newcommand{\be}{\begin{equation}}
\newcommand{\ee}{\end{equation}}
\newcommand{\bea}{\begin{eqnarray}}
\newcommand{\eea}{\end{eqnarray}}
\newcommand{\da}{\dagger}
\newcommand{\dg}[1]{\mbox{${#1}^{\dagger}$}}
\newcommand{\hlf}{\mbox{$1\over2$}}
\newcommand{\lfrac}[2]{\mbox{${#1}\over{#2}$}}
\newcommand{\scsz}[1]{\mbox{\scriptsize ${#1}$}}
\newcommand{\tsz}[1]{\mbox{\tiny ${#1}$}}

\begin{center}

\Large{\bf SCHR\"ODINGER EQUATIONS WITH TIME-DEPENDENT
${\mathbf P^2}$ AND ${\mathbf X^2}$ TERMS}

\vspace{0.5in}

MICHAEL MARTIN NIETO \\ 

{\it Theoretical Division (MS-B285), Los Alamos National Laboratory,\\
University of California, 
Los Alamos, New Mexico 87545, U.S.A. \\
mmn@lanl.gov}

\vspace{0.5in}

D. RODNEY TRUAX \\
{\it Department of Chemistry, University of Calgary,\\
Calgary, Alberta T2N 1N4, Canada \\
truax@ucalgary.ca}

\vskip 20pt
\today

\end{center}

\vspace{0.3in}

\baselineskip=.33in

\noindent 
We present some general results for the time-dependent mass 
Hamiltonian problem with ${H}=-\lfrac{1}{2}e^{-2\nu(t)}\partial_{xx} 
+h^{(2)}(t)e^{2\nu(t)}x^2$, where $\nu(t)$ is a continuous function 
of $t$.  This Hamiltonian corresponds to a 
time-dependent mass ($TM$) Schr\"odinger equation with the 
restriction that there are only $P^2$ and $X^2$ terms.   
We give the specific transformations to a
different quadratic Schr\"odinger($TQ$) equation and to a different 
time-dependent oscillator ($TO$) equation.  
For each Schr\"odinger system, 
we give the Lie algebra of space-time symmetries and $(x,t)$  
representations for number states, coherent states, and squeezed states.
These general results include earlier work as special cases.

\vspace{.25in}


\newpage

\baselineskip=.33in

\section{Introduction}

Recently \cite{paperI,paperII}, we  studied 
general time-dependent quadratic ($TQ$) Schr\"odinger equations
and Hamiltonians 
\begin{eqnarray}
S_1\Phi(x,t) & = & \{-\left[1+k(t)\right]P^2+2T+h(t)D+
                       g(t)P    \nonumber\\
               &   & \hspace{1.0cm}-2h^{(2)}(t)X^2-2h^{(1)}(t)X
                    -2h^{(0)}(t)I\}\Phi(x,t) = 0, \nonumber
\eea
\begin{eqnarray}
 & T=i\partial_{t},~~~~~P=-i\partial_x,~~~~~X=x,~~~~~I=1, & 
\nonumber \\*[1mm]
 & P^2=-\partial_{xx},~~~~~X^2=x^2,~~~~~D=\lfrac{1}{2}\left(XP+PX\right) 
= -ix\partial_x-i/2. & \label{e:pre12} 
\end{eqnarray}
\bea
H_1(x,t) & = & \frac{[1+k(t)]}{2}P^2-\frac{h(t)}{2}D
                       -\frac{g(t)}{2}P    
               +h^{(2)}(t)X^2 + h^{(1)}(t)X + h^{(0)}(t)I, \label{genH}
\end{eqnarray}
where $\{k(t),h(t),g(t),h^{(2)}(t),h^{(1)}(t),h^{(0)}(t)\}$ are
arbitrary functions of time.  We showed in Ref. \cite{paperI} that 
these are related to time-dependent mass ($TM$) equations
\begin{eqnarray}
{S}_2\Theta(x,t) & = & \left\{-f(t)P^2+2T -2{f}^{(2)}
(t)X^2 -2{f}^{(1)}(t)X-2{f}^{(0)}(t)I\right\}\Theta(x,t)  =  0  
\label{e:pre8}
\end{eqnarray}
by unitary transformations.  
We also demonstrated that the $TM$ equations are in turn related 
to time-dependent oscillator ($TO$) equations of the form 
\begin{equation}
{S}_3\Psi(x,t') = \{-P^2+2T'-2g^{(2)}(t')X^2-2g^{(1)}(t')X-
2g^{(0)}(t')I\}\Psi(x,t')=0,  \label{e:pre1}
\end{equation}
by time transformations $t(t')$.
In Ref. \cite{paperII} we obtained the isomorphic
Schr\"odinger algebras for these three types of systems and 
the general characteristics of the number states, coherent states, 
and squeezed states.  

In the present paper we will show in detail how to obtain 
the number states, coherent states, 
and squeezed states for a certain subclass of these systems: 
specifically those which come from $TM$ Hamiltonians with only 
time-dependent $P^2$ and $X^2$ terms. Some examples of 
this $TM$ subclass of equations have been partially discussed 
in the literature \cite{ek1}-\cite{mc}. 
Further, elsewhere we have applied our formalism to two specific  
Hamiltonians, obtaining very general new results
\cite{ex1,ex2,usfirst}. 
Those examples are special cases of the more general results given
here. 

We begin with Section 2, where 
we describe this subclass of $TM$ Schr\"odinger 
equations and show the mappings from $TQ$ Schr\"odinger 
equations and to $TO$ Schr\"odinger equations.  
The main reason for mapping to $TO$ equations is that we have an 
algorithm for computing the operators that form a basis for the Lie, 
space-time symmetry algebra for $TO$ equations \cite{drt1}-\cite{nt1}.  
This algebra is called a Schr\"odinger algebra, and has an oscillator 
subalgebra, $os(1)$.  

In Section 3 we construct the algebra for the $TO$ equation.  
Then, using the mappings discussed 
above, we construct the analogous $os(1)$ 
operators for the $TM$ equation and 
the $TQ$ equation.  Using the structure of the 
oscillator subalgebra, $os(1)$, of the $TO$ Schr\"odinger algebra, 
in Section 4 we  
construct a set of number states that are solutions to 
the $TO$ Schr\"odinger equation.  We then proceed and do the 
same for the $TM$ and $TQ$ Schr\"odinger equations.

Next, for the $TO$ equation, 
we can formulate displacement-operator coherent-state (DOCS) 
wave functions  
from the extremal number state and the ladder operators of the $os(1)$ 
algebra.  Employing the transformations mentioned 
above, we can then transform  the $TO$ coherent 
states into the corresponding coherent states 
for the $TM$ and $TQ$ Schr\"odinger equations.  
This is done in Section 5.  The analogous program is done for the
DOSS (squeezed-states)  in Section 6.  


\section{The Subclass of Equations}

The subclass of TM equations of interest to us have the form 
\begin{eqnarray}
\hat{S}_2\hat{\Theta}(x,t) & = & \left\{-2\hat{H}_2+2T\right\}
\hat{\Theta}(x,t)=0,\label{genS2}\\
\hat{H}_2 & = & \lfrac{1}{2}e^{-2\nu(t)}P^2
+h^{(2)}(t)e^{2\nu(t)}X^2,
\label{genH2}
\end{eqnarray}
where $\nu(t)$ is a real function of $t$.
This class of Schr\"odinger equations can be obtained from the subclass 
of TQ equations
\begin{eqnarray}
S_1\Phi(x,t)&=&\left\{-2H_1+2T\right\}\Phi(x,t)=0,\label{genS1} \\
H_1&=&\lfrac{1}{2}P^2-\lfrac{1}{2}h(t)D+h^{(2)}(t)X^2, \label{genH1}
\end{eqnarray}
by the unitary transformation
\begin{eqnarray}
&R(0,\nu(t),0)=\exp\left\{i\nu(t) D\right\},& \label{tr1}  \\
&\nu(t)=-\lfrac{1}{2}\int_{t_o}^{t}ds\,h(s).&  \label{nudef}
\end{eqnarray}
The Hamiltonians of Refs. \cite{ex1,ex2} are examples of this type of $TM$ 
Schr\"odinger equation, which were analyzed in detail there. 

In order to solve this problem using algebraic methods, we 
transform 
the $TM$ Schr\"odinger equation (\ref{genS2}) into a 
$TO$ Schr\"odinger equation
\begin{equation}
S_3\Psi(x,t')=\left\{-2H_3+2T'\right\}\Psi(x,t')=0, 
   \label{genS3}  
\end{equation}
by a specific change in the ``time" variable, $t\rightarrow t'$, and   
with $T'=i\partial_{t'}$.  The resulting $TO$ Hamiltonian is 
\begin{eqnarray}
& H_3=\lfrac{1}{2}P^2 +g^{(2)}(t')X^2, &  \label{genH3} \\
& g^{(2)}(t')=\check{h}^{(2)}(t')e^{4\check{\nu}(t')}, &
    \label{e:genH3.11}  \\
& \check{h}^{(2)}(t')=(h^{2)}\circ t)(t'),~~~~~
        \check{\nu}(t')=(\nu\circ t)(t'),&
\label{tr4}
\end{eqnarray}
all time dependence now being in the single function $g^{(2)}(t')$.
  The operators $X$, $X^2$, $P$, $P^2$, and $D$ 
and  their commutation relations are given in Eq. (\ref{e:pre12}).

According to Eq. (I-67), the time transformation is given by 
\begin{equation}
t'-t_o' =\int_{t_o}^t ds\,e^{-2\nu(s)}.\label{e:tr8}
\end{equation}
We assume that the mapping $t'(t)$, described in Eq. (\ref{e:tr8}), has 
an inverse $t(t')$.  This was certainly true for the Hamiltonians of 
Refs. \cite{ex1,ex2}


\section{The Algebras for our Problem}

\subsection{
The $TO$ Schr\"odinger algebra, $({\cal SA})_1^c$, 
and its $os(1)$ Subalgebra}

Combining Eqs. (\ref{genS3}) and (\ref{genH3}), we obtain the equation 
\begin{equation}
S_3\Psi(x,t')=\left\{-P^2+2T'-2g^{(2)}(t')X^2\right\}\Psi(x,t')=0.
\label{genS3a}
\end{equation}
It is known {\cite{drt2,nt1}} that this equation has the Schr\"odinger 
algebra $({\cal SA})_1^c$ as its Lie algebra of space-time symmetries.  
This algebra is spanned by six operators.   

Three of the operators form an $su(1,1)$ subalgebra,  
\begin{eqnarray}
M_{3-} & = & \phi_1T' -\lfrac{1}{2}\dot{\phi}_1 D +\lfrac{1}{4}
\ddot{\phi}_1X^2,\nonumber\\*[1mm]
M_{3+} & = & \phi_2T' -\lfrac{1}{2}\dot{\phi}_2 D +\lfrac{1}{4}
\ddot{\phi}_2X^2,\nonumber\\*[1mm]
M_3 & = & \phi_3T' -\lfrac{1}{2}\dot{\phi}_3 D +\lfrac{1}{4}
\ddot{\phi}_3X^2.   \label{suop3.1}
\end{eqnarray}
[The `dot' over a symbol will be reserved for differentiation by $t'$.] 
Three others form an Heisenberg-Weyl subalgebra, $w_1^c$,  
\begin{eqnarray}
 & J_{3-} = i\left\{\xi P - \dot{\xi} X\right\},
~~~J_{3+}=i\left\{-\bar{\xi} P + 
\dot{\bar{\xi}} X\right\}.~~~I=1, & \label{op3.1}
\end{eqnarray}
 
The function, $\xi$, and its complex conjugate, $\bar{\xi}$ are 
solutions to the  differential equation 
\begin{equation}
\ddot{\gamma}+2g^{(2)}(t')\gamma=0,\label{ode3.1}
\end{equation}
where, as indicated, $\dot{\gamma}\equiv d\gamma/dt'$.  
These solutions satisfy the Wronskian condition {\cite{drt2,aec1}}
\begin{equation}
W(\xi,\bar{\xi})=\xi\dot{\bar{\xi}}-\dot{\xi}\bar{\xi}=-i.\label{cwron3}
\end{equation}
The functions, $\phi_j$, $j=1,2,3$, are defined as 
\begin{equation}
\phi_1(t')=\xi^2(t'),~~~~\phi_2(t')=\bar{\xi}^2(t'),~~~~\phi_3(t') 
= 2\xi(t')\bar{\xi}(t').\label{phi}
\end{equation}

The $su(1,1)$ operators satisfy the commutation relations
\begin{equation}
[M_{3+},M_{3-}]=-M_3,~~~~[M_3,M_{3\pm}]=\pm 2M_{3\pm}.\label{sucom3}
\end{equation}
The $w_1^c$ operators satisfy the nonzero commutation relation 
\begin{equation}
[J_{3-},J_{3+}]=I.\label{wcom3}
\end{equation}
The remaining commutation relations are 
\begin{eqnarray}
 & [M_{3-},J_{3-}]=0,~~~[M_{3+},J_{3-}]=+J_{3+},~~~[M_3,J_{3-}]=-J_{3-}, &
\nonumber\\*[1mm]
 & [M_{3-},J_{3+}]=-J_{3-},~~~[M_{3+},J_{3+}]=0,~~~[M_3,J_{3+}]=+J_{3+}. & 
\label{swcomm3}
\end{eqnarray}

In order to find solutions to Eq. (\ref{genS3a}), we need not consider 
the entire algebra $({\cal SA})_1^c$, only the 
$os(1)$ subalgebra consisting 
of the operators $\{M_3,J_{3\pm},I\}$.  From the above commutation 
relations, we see that these operators close to give the nonzero 
commutators 
\begin{equation}
[M_3,J_{3\pm}]=\pm J_{3\pm},~~~~[J_{3-},J_{3+}]=I.\label{oscom3}
\end{equation}
 Eq. (\ref{oscom3}) has 
the structure of an $os(1)$ oscillator algebra.  
Closure of the oscillator subalgebra follows from the Wronskian (\ref{cwron3})
and the relationships
\begin{eqnarray}
 & i\xi = \phi_3\dot{\xi}-\lfrac{1}{2}\dot{\phi}_3\xi, & \\
 & i\dot{\xi} =
\frac{d}{dt'} \left(\phi_3\dot{\xi}-
\lfrac{1}{2}\dot{\phi}_3\xi\right)
=
\phi_3\ddot{\xi}+\lfrac{1}{2}\dot{\phi}_3\dot{\xi}-
\lfrac{1}{2}\ddot{\phi}_3\xi, & 
\label{oscomfns3} 
\end{eqnarray}
Indeed, this subset of operators and commutation relations may be viewed 
as a generalization of the usual oscillator algebra.

Although we can 
construct the full Schr\"odinger algebras for each subclass of TO
(\ref{genS3}), TM (\ref{genS2}), and TQ (\ref{genS1}) equations, 
explicitly fully written in Eq. (\ref{genS3a}) and below, 
\begin{eqnarray}
\hat{S}_2\hat{\Theta}(x,t) & = & \left\{-e^{-2\nu}P^2+2T-2e^{2\nu}X^2\right\}
\hat{\Theta}(x,t)=0,\label{genS2a}\\*[2mm]
S_1\Phi(x,t) & = & \left\{-P^2+2T+h(t)D-2h^{(2)}(t)X^2\right\}\Phi(x,t)=0,
\label{genS1a}
\end{eqnarray}
we shall henceforth consider only the $os(1)$ subalgebras 
for all three systems.


\subsection{
The $os(1)$ algebra for $TM$ Schr\"odinger Equations}

The operators that span the $os(1)$ 
algebra for Eq. (\ref{genS2}) are 
\begin{eqnarray}
 & \hat{M}_2 =\hat{\phi}_3e^{2\nu}T-\lfrac{1}{2}\hat{\dot{\phi}}_3D+
\lfrac{1}{4}\hat{\ddot{\phi}}_3X^2, & \label{suop2}\\*[1mm]
 & \hat{J}_{2-} = i\left\{\hat{\xi}P - \hat{\dot{\xi}}X\right\},~~~
\hat{J}_{2+} = i\left\{-\hat{\bar{\xi}}
P + \hat{\dot{\bar{\xi}}}X\right\},~~~I=1,\label{hwop2}  \\
& \hat{\xi}(t)=(\xi\circ t')(t),~~~\hat{\dot{\xi}}(t)=(\dot{\xi}\circ t')
(t), & \nonumber\\*[1mm]
 & \hat{\phi}_3(t)=(\phi_3\circ t')(t),~~~\hat{\dot{\phi}}_3(t)=
(\dot{\phi}_3\circ t')(t),~~~\hat{\ddot{\phi}}_3(t)=(\ddot{\phi}_3
\circ t')(t). & \label{hatfns2}
\end{eqnarray}
(See Eqs. (31) and (36) of Ref. \cite{paperII}.)
The functions $\hat{\xi}$ and $\hat{\bar{\xi}}$ satisfy the analogue of the
Wronskian (\ref{cwron3})
\begin{equation}
\hat{\xi}\hat{\dot{\bar{\xi}}}-\hat{\dot{\xi}}\hat{\bar{\xi}}=-i.
\label{tmwron2}
\end{equation}
It is important to note that, in general,
\begin{equation}
 \hat{\dot{\xi}}\equiv \frac{d}{dt'}\hat{\xi}
= \frac{dt}{dt'}\frac{d \hat{\xi}}{dt} \ne \frac{d \hat{\xi}}{dt}.     
\label{derivs}
\end{equation}

The operators (\ref{suop2}) and (\ref{hwop2}) satisfy the nonzero 
commutator brackets
\begin{equation}
[\hat{M}_2,\hat{J}_{2\pm}]=\pm\hat{J}_{2\pm},~~~[\hat{J}_{2-},
\hat{J}_{2+}]=I.\label{oscom2}
\end{equation}
The structure of this algebra is isomorphic to $os(1)$.  Because of
this  we refer to it as an $os(1)$ algebra also.   
Closure of the commutators in (\ref{oscom2}), is the result of Eq. 
(\ref{tmwron2}) and 
\begin{eqnarray}
 & i\hat{\xi} = \hat{\phi}_3\hat{\dot{\xi}}
       -\lfrac{1}{2}\hat{\dot{\phi}}_3\hat{\xi}, & \\
 & i\hat{\dot{\xi}} = \hat{\phi}_3\hat{\ddot{\xi}}+\lfrac{1}{2}
\hat{\dot{\phi}}_3\hat{\dot{\xi}}
-\lfrac{1}{2}\hat{\ddot{\phi}}_3\hat{\xi}. & \label{oscomfns2} 
\end{eqnarray}

To obtain  Eq. (\ref{oscomfns2}), 
as well as the earlier Eq. (\ref{oscomfns3}),
it is helpful to realize that
\begin{eqnarray}
&\ddot{\xi}=-2g^{(2)}(t')\xi \Rightarrow \hat{\ddot{\xi}}=-2h^{(2)}(t)
      e^{4\nu}\hat{\xi},&  
  \label{oscomfns3.1} \\
&\ddot{\phi}_3=-4g^{(2)}(t')\phi_3+4\dot{\xi}\dot{\bar{\xi}} \Rightarrow 
\hat{\ddot{\phi}}_3=-4h^{(2)}(t)e^{4\nu}\hat{\phi}_3+4\hat{\dot{\xi}}
\hat{\dot{\bar{\xi}}}.&
\label{oscomfns2.1}
\end{eqnarray}


\subsection{
The $os(1)$ algebra for $TQ$ Schr\"odinger Equations}
 
The following operators are Lie 
symmetries for the $TQ$ Schr\"odinger equation (\ref{genS1}):
\begin{eqnarray}
 & J_{1-} = i\left\{\Xi_PP-\Xi_XX\right\},~~~~~~~J_{1+} = i\left\{
-\bar{\Xi}_PP+\bar{\Xi}_XX\right\},~~~~~~~I=1, \nonumber\\*[1mm]
 & M_1 = C_{3,T}T - C_{3,D}D - C_{3,X^2}X^2, & \label{osop1.1}  \\
&\Xi_P(t)=\hat{\xi}(t)e^{\nu},~~~~~~~~\Xi_X(t)=\hat{\xi}(t)e^{-\nu}.&
\label{opfns1.1}
\end{eqnarray}
(See Eqs. (40)-(42) and (45)-(52) of Ref. \cite{paperII} 
with $\kappa=0$.) 
The coefficients, $C_{3,T}$, $C_{3,D}$, and $C_{3,X^2}$ are 
\begin{eqnarray}
C_{3,T} &=& \hat{\phi}_3(t)e^{2\nu}=2\Xi_P\bar{\Xi}_P, 
        \label{opfns1.4}  \\
C_{3,D} & = & -\lfrac{1}{2}h(t)\hat{\phi}_3(t)e^{2\nu}+\lfrac{1}{2}
\hat{\dot{\phi}}_3
  =  -\lfrac{1}{2}h(t)C_{3,T}+\Xi_P\bar{\Xi}_X+\Xi_X\bar{\Xi}_P, 
\label{opfns1.8}  \\
C_{3,X^2} &=& -\lfrac{1}{4}\hat{\ddot{\phi}}_3e^{-2\nu} 
=h^{(2)}(t) C_{3,T}-\Xi_X\bar{\Xi}_X. \label{opfns1.12}
\end{eqnarray}
The second equality of Eq. (\ref{opfns1.12}) 
follows from Eq. (\ref{oscomfns2.1}).
 
The $TQ$ symmetry operators satisfy the commutation relations 
\begin{equation}
[M_1,J_{1\pm}]=\pm J_{1\pm},~~~[J_{1-},J_{1+}]=I.\label{oscom1}
\end{equation}
This algebra is also isomorphic to the $os(1)$ and we shall refer to 
it as the $TQ$ oscillator algebra.  Closure of the $TQ$ commutator 
brackets in Eq. (\ref{oscom1}) follows from 
\begin{eqnarray}
 & i\Xi_P = C_{3,T}\left(\Xi_X-\lfrac{1}{2}h(t)\Xi_P\right) - 
C_{3,D}\Xi_P, & \nonumber\\*[1mm]
 & i\Xi_X = C_{3,T}\left(-2h^{(2)}(t)\Xi_P+\lfrac{1}{2}h(t)\Xi_X\right)
+C_{3,D}\Xi_X+2C_{3,X^2}\Xi_P.\label{oscom1fns}
\end{eqnarray}
The corresponding relationships for complex conjugates can be obtained by 
taking the complex conjugate of each relationship.  

Each of the three $os(1)$ algebras has a Casimir operator
\begin{equation}
{\bf C}_j = J_{j+}J_{j-}-M_jI,~~~~j=1,\hat{2},3, \label{casimir1}
\end{equation}
where $\hat{2}$ indicates that all operators with the subscript 2 should 
receive a hat.  We refer the reader to Eqs. 
(62), (64), and (65) of Ref. \cite{paperII} 
for their relationships to the corresponding Schr\"odinger operators. 

Next, we shall exploit the $os(1)$ algebraic structure to obtain number 
states.   
      

\section{Number States}

\subsection{
$\Psi_n(x,t')$ for $TO$ Systems}

It is possible to construct a set of number 
states by first solving a first-order partial differential equation 
for $\Psi_n(x,t')$ in terms of $\psi_n(x)$,
and then solving 
a first-order ordinary differential equation 
for the extremal state wave function, $\Psi_0(x,t')$. 
(See Section 4.2 of \cite{paperII} on representation theory of 
oscillator algebras.)  It is  important to reemphasize
that these number states are, in general, not eigenstates 
of any Hamiltonian.  As a consequence, $\Psi_0(x,t')$, does not 
necessarily represent a ground state.

With the above we have 
\begin{eqnarray}
 & M_3\Psi_n(x,t')=\left(n+\lfrac{1}{2}\right)\Psi_n(x,t'),~~~~
{\bf C}_3\Psi_n(x,t')=-\lfrac{1}{2}\Psi_n(x,t'), & \label{osrep3.1}\\*[1mm]
 & J_{3+}\Psi_n(x,t')=\sqrt{n+1}\Psi_{n+1}(x,t'),~~~~J_{3-}\Psi_n(x,t')
=\sqrt{n}\Psi_{n-1}(x,t'), & \label{osrep3.2}
\end{eqnarray}
where $n\in {\bf Z}_0^+$ is the set of integers $\ge 0$.  
The constraint that the spectrum of the operator $M_3$, ${\rm Sp}(M_3)$, 
be bounded below is that 
\begin{equation}
J_{3-}\Psi_0(x,t')=0.\label{osrep3.4}
\end{equation}

Using Eqs.  (\ref{suop3.1}) and (\ref{osrep3.1}), 
with Eqs. (6) and (7) of \cite{paperI}, we 
obtain a first-order partial differential equation for $\Psi_n$ which 
can be integrated by the method of characteristics to yield a solution 
of the type {\cite{drt2,zt}}
\begin{equation}
\Psi_n(x,t') = \exp{\left\{\lfrac{i}{4}\frac{\dot{\phi}_3}{\phi_3}x^2
\right\}}\psi_n\left(\frac{x}{\phi_3^{1/2}}\right)\phi_3^{-1/4}
\left(\frac{\bar{\xi}}{\xi}\right)^{\lfrac{1}{2}\left(n+\lfrac{1}{2}
\right)}.\label{ns3.1}
\end{equation}
For $n=0$, Eq. (\ref{osrep3.4}) with $\Psi_0$ given by Eq. (\ref{ns3.1}) 
yields a first-order ordinary differential equation for $\psi_0$ whose 
solution is 
\begin{equation}
\psi_0\left(\frac{x}{\phi_3^{1/2}}\right)=N\exp\left(-\frac{x^2}{2\phi_3}
\right),\label{ns3.4}
\end{equation}
where $N$ is an integration constant.  Thus, the normalized extremal state is 
\begin{equation}
\Psi_0(x,t')=\exp{\left\{\frac{i}{4}\frac{\dot{\phi}_3}{\phi_3}x^2\right\}}
\exp{\left(-\frac{x^2}{2\phi_3}\right)}(\pi\phi_3)^{-1/4}
\left(\frac{\bar{\xi}}{\xi}\right)^{\lfrac{1}{4}}.\label{ns3.8}
\end{equation}

The higher-order states are obtained by the repeated application of 
the raising operator $J_{3+}$.  The normalized higher-order states are
\begin{eqnarray}
\Psi_n(x,t') & = & \sqrt{\frac{1}{n!}}\left(J_{3+}\right)^n\Psi_0(x,t')
\nonumber\\*[1mm]
  & =  & \sqrt{\frac{1}{n!2^n}}\exp{\left\{\lfrac{i}{4}
\frac{\dot{\phi}_3}{\phi_3}x^2\right\}}H_n\left(\frac{x}{\phi_3^{1/2}}
\right)\exp{\left(-\frac{x^2}{2\phi_3}\right)}
(\pi\phi_3)^{-1/4}\left(\frac{\bar{\xi}}{\xi}
\right)^{\lfrac{1}{2}\left(n+\lfrac{1}{2}\right)}, \label{ns3.12}
\end{eqnarray}
where $H_n(x/\phi_3^{1/2})$ is a Hermite polynomial of order $n$.


\subsection{
$\hat{\Theta}_n(x,t)$ for $TM$ Systems}

An analogous set of relationships to Eqs. (\ref{osrep3.1}) and 
(\ref{osrep3.2}) can be written for $TM$-$os(1)$ representation spaces
of Eq. (\ref{genH2}).
They are
\begin{eqnarray}
 & \hat{M}_2\hat{\Theta}_n(x,t)=\left(n+\lfrac{1}{2}\right)
\hat{\Theta}_n(x,t),~~~~\hat{{\bf C}}_2\hat{\Theta}_n(x,t)=-\lfrac{1}{2}
\hat{\Theta}_n(x,t), & \label{osrep2.1}\\*[1mm]
 & \hat{J}_{2+}\hat{\Theta}_n(x,t)=\sqrt{n+1}\hat{\Theta}_{n+1}(x,t),
~~~~\hat{J}_{2-}\hat{\Theta}_n(x,t)
=\sqrt{n}\hat{\Theta}_{n-1}(x,t), &\label{osrep2.2}
\end{eqnarray}  
with  $n\in {\bf Z}_0^+$.  
Also, we have the constraint on the extremal state
\begin{equation}
\hat{J}_{2-}\hat{\Theta}_0(x,t)=0.\label{osrep2.4}
\end{equation}

To obtain $\hat{\Theta}_n$, we could follow the same procedure as we 
did for $TO$ number states.  But it is more convenient to take the 
composition of $\Psi_n(x,t')$ with $t'(t)$.  Then
\begin{eqnarray}
& \hat{\Theta}_n(x,t)   =  \sqrt{\frac{1}{n!2^n}}\exp{\left\{\lfrac{i}{4}
\frac{\hat{\dot{\phi}}_3}{\hat{\phi}_3}x^2\right\}}H_n
\left(\frac{x}{\hat{\phi}_3^{1/2}}
\right)\exp{\left(-\frac{x^2}{2\hat{\phi}_3}\right)}
(\pi\hat{\phi}_3)^{-1/4}
\left(\frac{\hat{\bar{\xi}}}{\hat{\xi}}\right)^{\lfrac{1}{2}
\left(n+\lfrac{1}{2}\right)}, &  \label{ns2.1}
\end{eqnarray}
where $\hat{\xi}$, $\hat{\phi}_3$, and $\hat{\dot{\phi}}_3$ are defined 
in Eq. (\ref{hatfns2}).  
Setting $n=0$ yields the extremal state, $\hat{\Theta}_0(x,t)$. 


\subsection{
$\Phi_n(x,t)$ for $TQ$ Systems}

For $TQ$  systems, we have 
\begin{eqnarray}
 & M_1\Phi_n(x,t)=\left(n+\lfrac{1}{2}\right)\Phi_n(x,t),~~~~
{\bf C}_1\Phi_n(x,t)=-\lfrac{1}{2}\Phi_n(x,t), & \label{osrep1.1}
\\*[1mm]
 & J_{1+}\Phi_n(x,t)=\sqrt{n+1}\Phi_{n+1}(x,t),~~~~J_{1-}\Phi_n(x,t)
=\sqrt{n}\Phi_{n-1}(x,t), & \label{osrep1.2}
\end{eqnarray}
with  $n\in {\bf Z}_0^+$. 
This is subject to the constraint
\begin{equation}
J_{1-}\Phi_0(x,t)=0.\label{osrep1.4}
\end{equation}

The wave functions, $\Phi_n(x,t)$, satisfying 
these relationships can be conveniently obtained as 
\begin{eqnarray}
& \Phi_n(x,t)  =  \exp{[-i\nu D]}\hat{\Theta}_n(x,t)
   =  e^{-\nu/2}\hat{\Theta}_n(xe^{-\nu},t), &  \label{ns1.1}
\end{eqnarray}
for $n\in {\bf Z}_0^+$.  According to Eq. (\ref{ns2.1}), the number state, 
$\Phi_n(x,t)$, is 
\begin{eqnarray}
& &\Phi_n(x,t)  =  \sqrt{\lfrac{1}{n!}}\left(J_{1+}\right)^n\Phi_0(x,t)
\nonumber\\*[1mm]
&  & ~~ =  \sqrt{\frac{1}{n!2^n}}\exp{\left(\lfrac{i}{4}
\frac{\hat{\dot{\phi}}_3}{C_{3,T}}x^2\right)}H_n\left(\frac{x}
{C_{3,T}^{1/2}}\right)\exp{\left(-\frac{x^2}{2C_{3,T}}\right)}
(\pi C_{3,T})^{-1/4}\left(\frac{\bar{\Xi}_P}{\Xi_P}
\right)^{\lfrac{1}{2}\left(n+\lfrac{1}{2}\right)},
\label{ns1.4}
\end{eqnarray}
where $C_{3,T}$, which is real and positive, is given by Eq. 
(\ref{opfns1.4}).  The functions $\Xi_P$ and $\Xi_X$ are given by Eq.
(\ref{opfns1.1}).  The extremal state, $\Phi_0(x,t)$, is obtained by 
setting $n=0$.


\section{Coherent States}

Now that we have computed the extremal states for the $TO$, 
$TM$, and $TQ$ systems, we can compute explicit $(x,t)$ 
representations of coherent-state and 
squeezed-state wave functions for each of the three classes of systems.  
Although the wave functions themselves are not needed 
for the calculation of expectation values, they are independently 
of interest.  In this section, we  calculate coherent-state 
wave functions using a group theoretic method.  
Table 1 defines the CS parameters $\alpha$ that will be used. 
It is obtained 
from Eq. (83) and Table 2 of \cite{paperII}, with $\kappa=\mu=0$ and 
$F_P=F_X=0$.  [Here and in the following a superscript symbol $^o$ 
denotes that the function is evaluated at initial time.]

\begin{center}
\begin{tabular}{|c|c|c|c|}
\multicolumn{4}{l}{Table 1. Generic functions and their values according 
to class used}\\
\multicolumn{4}{l}{to obtain the CS parameters 
$\alpha=i\left(G_P^op_o-G_X^ox_o\right)$.}\\*[.2cm]
\hline\hline
\multicolumn{1}{|c|}{} & \multicolumn{1}{c|}{} & 
\multicolumn{1}{c|}{} & 
\multicolumn{1}{c|}{} \\*[-.2cm]
\multicolumn{1}{|c|}{Function} & \multicolumn{1}{c|}{$TO$}
& \multicolumn{1}{c|}{$TM$} & \multicolumn{1}{c|}{$TQ$}\\*[.2cm]\hline
  & {\hspace{3cm}} & {\hspace{3cm}} & {\hspace{3cm}} \\*[-.2cm]
$G_P^o$ & $\xi^o$  & $\hat{\xi}^o$ & 
$\Xi_P^o=\hat{\xi}^o$\\*[.2cm]\hline
  &  &  & \\*[-.2cm]
$G_X^o$ & $\dot{\xi}^o$ & $\hat{\dot{\xi}}^o$ & 
$\Xi_X^o=\hat{\dot{\xi}}^o$ 
\\*[.3cm]
\hline\hline
\end{tabular}
\end{center}


\subsection{
$TO$ Coherent-State Wave Functions}

For $TO$ systems we have 
\begin{eqnarray}
\Psi_{\alpha}(x,t') & = & D(\alpha)\Psi_0(x,t')=
\exp{\left(-\lfrac{1}{2}|\alpha|^2I\right)}\exp{
\left(\alpha J_{3+}\right)}\Psi_0(x,t')\label{cs3.1}\\*[1mm]
  & = & \exp{\left(-\lfrac{1}{2}|\alpha|^2\right)}\exp{\left[i\dot{\bar{\xi}}
\left(x\alpha-\lfrac{1}{2}\bar{\xi}\alpha^2\right)\right]}
\Psi_0(x-\alpha\bar{\xi},t')\label{cs3.2}\\*[1mm]
  & = & \left(\pi\phi_3\right)^{-1/4}\left(\frac{\bar{\xi}}{\xi}
\right)^{\lfrac{1}{4}}\exp\left\{-\frac{1}{2\phi_3}\left[x-X_3^+(\alpha)
\right]^2\right\}\nonumber\\
  &   & \times\exp\left\{
i\left[\frac{1}{4}\frac{\dot{\phi}_3}{\phi_3}x^2
+\frac{1}{\phi_3}\left(x-\lfrac{1}{2}X_3^+(\alpha)\right)X_3^-(\alpha)
\right]\right\},
\label{cs3.3}  \\
X_3^{+}(\alpha) &=& \alpha\bar{\xi}+\bar{\alpha}\xi
= i(\xi^o\bar{\xi}-\bar{\xi}^o\xi)p_o
-i(\dot{\xi}^o\bar{\xi}-\dot{\bar{\xi}}^{\,o}\xi)x_o,
\nonumber \\
X_3^{-}(\alpha) &=&-i( \alpha\bar{\xi}-\bar{\alpha}\xi)
= (\xi^o\bar{\xi}+\bar{\xi}^o\xi)p_o
-(\dot{\xi}^o\bar{\xi}+\dot{\bar{\xi}}^{\,o}\xi)x_o.
\label{csfns1.1}
\end{eqnarray} 
To obtain Eq. (\ref{cs3.1}) we used the standard 
Baker-Campbell-Hausdorff (BCH) relation 
{\cite{jk,amp}} for the operator 
$D(\alpha)=\exp[\alpha J_{3+} - \bar{\alpha}J_{3-}]$
and the fact that $J_{3-}$ 
annihilates the extremal state.  For Eq. (\ref{cs3.2}), see 
Ref. {\cite{wm1}}. To obtain Eq. (\ref{cs3.3}), we  employed 
Eqs. (\ref{ns3.8}) and (\ref{cwron3}). 
The explicit definition of $\alpha$ can be found from Table 1. 

Note that both the $X_3^{\pm}(\alpha)$ 
defined in Eq. (\ref{csfns1.1}) are real. 
Therefore, 
the final expression,  Eq. (\ref{cs3.3}),   has been 
arranged in such a way that the argument of the first exponential 
is real while the argument of the second exponential is pure imaginary.


\subsection{
$TM$ Coherent-State Wave Functions}

In the case of $TM$ coherent states, we take the composition of 
$\Psi_{\alpha}(x,t')$ and $t'(t)$ to obtain
\begin{eqnarray}
\hat{\Theta}_{\alpha}(x,t) & = & \left(\pi\hat{\phi}_3\right)^{-1/4}
\left(\frac{\hat{\bar{\xi}}}{\hat{\xi}}\right)^{\lfrac{1}{4}}
\exp\left\{-\frac{1}{2\hat{\phi}_3}\left[x-\hat{X}_2^+(\alpha)
\right]^2\right\}\nonumber\\
 &  & \times\exp\left\{
i\left[\frac{1}{4}\frac{\hat{\dot{\phi}}_3}{\hat{\phi}_3}
x^2+\frac{1}{\hat{\phi}_3}
\left(x-\lfrac{1}{2}\hat{X}_2^+(\alpha)\right)
\hat{X}_2^-(\alpha)\right]\right\},
\label{cs2.1}  \\
\hat{X}_2^{+}(\alpha) &=& 
\alpha\hat{\bar{\xi}}+\bar{\alpha}\hat{\xi} 
= i(\hat{\xi}^o\hat{\bar{\xi}}-\hat{\bar{\xi}}^o\hat{\xi})p_o
-i(\hat{\dot{\xi}}^o\hat{\bar{\xi}}
               -\hat{\dot{\bar{\xi}}}^{\,o}\hat{\xi})x_o,
\nonumber   \\
\hat{X}_2^{-}(\alpha) &=& 
-i(\alpha\hat{\bar{\xi}}-\bar{\alpha}\hat{\xi} 
= (\hat{\xi}^o\hat{\bar{\xi}}+\hat{\bar{\xi}}^o\hat{\xi})p_o
-(\hat{\dot{\xi}}^o\hat{\bar{\xi}}
               +\hat{\dot{\bar{\xi}}}^{\,o}\hat{\xi})x_o.
\label{csfns2.1}
\end{eqnarray}  
See Table 1 for the explicit definition of $\alpha$.  
Once more both expressions in Eq. (\ref{csfns2.1}) are real.


\subsection{
$TQ$ Coherent-State Wave Functions}

For $TQ$ systems, we calculate the coherent states from 
$TM$ coherent states.  
Starting from the definition of a $TQ$ coherent state, we have
\begin{eqnarray}
\Phi_{\alpha}(x,t) & = & \exp{\left(\alpha J_{1+}-\bar{\alpha}J_{1-}\right)}
\exp{\left(-i\nu D\right)}\hat{\Theta}_0(x,t)\label{cs1.1}\\*[1mm]
  & = & \exp{\left(-i\nu D\right)}\left[\exp{\left(i\nu D\right)}
\exp{\left(\alpha J_{1+}-\bar{\alpha}J_{1-}\right)}
\exp{\left(-i\nu D\right)}\right]\hat{\Theta}_0(x,t)\nonumber\\*[1mm]
  & = & \exp{\left(-i\nu D\right)}
            \left[\exp{\left(\alpha e^{i\nu D}J_{1+}e^{-i\nu D}
            -\bar{\alpha}e^{i\nu D}J_{1-}e^{-i\nu D}\right)}\right]
          \hat{\Theta}_0(x,t)\label{cs1.2}\\*[1mm]
  & = & \exp{\left(-i\nu D\right)}\exp{\left(\alpha \hat{J}_{2+}
-\bar{\alpha}\hat{J}_{2-}\right)}\hat{\Theta}_0(x,t).
\label{cs1.3}
\end{eqnarray}
(Further details on the transformation to Eq. (\ref{cs1.3}) can be 
found in the material adjoining  Eqs. (40) and (41) of \cite{paperII}.)  
Therefore, we obtain 
\begin{eqnarray}
\Phi_{\alpha}(x,t)  & = & \exp{\left(-i\nu D\right)}\hat{\Theta}_{\alpha}
(x,t) = \exp{\left(-\lfrac{1}{2}\nu\right)}
         \hat{\Theta}_{\alpha}(xe^{-\nu},t)  
\label{cs1.4}  \\
  & = & \left(\pi C_{3,T}\right)^{-1/4}\left(\frac{\bar{\Xi}_P}{\Xi_P}
\right)^{\lfrac{1}{4}}
\exp\left\{-\frac{1}{2C_{3,T}}\left[x-X_1^+(\alpha)\right]^2\right\}
   \nonumber\\
 &   & \times\exp\left\{
i\left[\frac{1}{4}\frac{\hat{\dot{\phi}}_3}{C_{3,T}}
x^2+\frac{1}{C_{3,T}}
\left(x-\lfrac{1}{2}X_1^+(\alpha)\right)X_1^-(\alpha)
\right]\right\},
\label{cs1.5}  \\ 
X_1^{+}(\alpha) &=& \alpha\bar{\Xi}_P+\bar{\alpha}\Xi_P 
= i(\Xi_P^o\bar{\Xi}_P-\bar{\Xi}_P^o\Xi_P)p_o
-i(\Xi_X^o\bar{\Xi}_P -\bar{\Xi}_X^o\Xi_P)x_o.
\nonumber  \\
X_1^{-}(\alpha) &=& -i(\alpha\bar{\Xi}_P-\bar{\alpha}\Xi_P) 
= (\Xi_P^o\bar{\Xi}_P+\bar{\Xi}_P^o\Xi_P)p_o
-(\Xi_X^o\bar{\Xi}_P +\bar{\Xi}_X^o\Xi_P)x_o.
\label{csfns1.4}
\end{eqnarray}
The explicit definitions of the $\Xi$
can be found in Table 1.Again the $X_1^{\pm}(\alpha)$ are both real.


\section{Squeezed States}

In this section, we derive expressions 
for the squeezed-state wave functions by solving a first-order partial 
differential equation. 

\subsection{
$TO$ Squeezed-State Wave Functions}

To calculate the squeezed state wave functions for a $TO$ system we write   
\bea
\Psi_{\alpha,z}(x,t) &=& D(\alpha)S(z)\Psi_0(x,t), \nonumber \\
S(z) &=& \exp{(z{\cal K}_+-\bar{z}{\cal K}_-)}
     =\exp{(\gamma_+{\cal K}_+)}\exp{(\gamma_3{\cal K}_3)}, 
       \exp{(\gamma_-{\cal K}_-)},    \nonumber \\
\gamma_- &=&-{{\bar{z}}\over{|z|}}\tanh{|z|},~~~~~
\gamma_+= {{z}\over{|z|}}\tanh{|z|},~~~~~
\gamma_3=-\ln{\left(\cosh{|z|}\right)},          \nonumber \\
\Psi_{\alpha,z}(x,t)&=&\exp{\left[\lfrac{1}{2}(\gamma_3-|\alpha|^2)\right]}
\exp{\left[(\alpha-\gamma_+\bar{\alpha})J_{3+}\right]}\exp{\left[\gamma_+
{\cal K}_+\right]}\Psi_0(x,t),\label{ss3.1}
\eea
where \{${\cal K}_{-}=\lfrac{1}{2}J_{3-}^2,~{\cal K}_{+}=\lfrac{1}{2}
J_{3+}^2,~{\cal K}_{3}=J_{3+}J_{3-}+\lfrac{1}{2}$\} satisfy the
$su(1,1)$ squeeze algebra 
and the last equality is obtained using standard methods 
Refs. {\cite{paperII,nt1,nt2}}.
The wave function (\ref{ss3.1}) satisfies the following eigenvalue equation
\begin{equation}
\left(J_{3-}-\gamma_+J_{3+}\right)\Psi_{\alpha,z}=(\alpha-\gamma_+
\bar{\alpha})\Psi_{\alpha,z}.\label{sspde3.1}
\end{equation}
The operators $J_{3-}$ and $J_{3+}$ are first-order partial differential 
operators of the form (\ref{op3.1}).  Hence, Eq. (\ref{sspde3.1}) is a 
first-order partial differential equation for $\Psi_{\alpha,z}$:
\begin{equation}
(\xi+\gamma_+\bar{\xi})\partial_x\Psi_{\alpha,z}-i(\dot{\xi}+\gamma_+
\dot{\bar{\xi}}x)\Psi_{\alpha,z}=(\alpha-\gamma_+\bar{\alpha})\Psi_{\alpha,z}.
\label{sspde3.4}
\end{equation}
We solve this equation by the method of characteristics {\cite{zt}} and 
obtain the following solution:
\begin{equation}
\Psi_{\alpha,z}(x,t')=b(t')\exp{\left[\frac{\alpha-\gamma_+\bar{\alpha}}
{\xi+\gamma_+\bar{\xi}}\,x+\lfrac{i}{2}\frac{\dot{\xi}+\gamma_+
\dot{\bar{\xi}}}{\xi+\gamma_+\bar{\xi}}\,x^2\right]}.\label{ss3.4}
\end{equation}

The function $b(t')$ is arbitrary.  To fix $b(t')$, we 
require that the wave function (\ref{ss3.4}) satisfy the Schr\"odinger 
equation 
$S_3\Psi_{\alpha,z}=\left\{\partial_{xx}+2i\partial_{t'}-2g^{(2)}(t')x^2
\right\}\Psi_{\alpha,z} =0$. 
[See Eq. (\ref{genS3a}) with 
$g^{(2)}(t')=\check{h}^{(2)}(t')e^{4\check{\nu}(t')}$.]   
This yields a first-order ordinary differential equation for $b(t')$:
\begin{equation}
\frac{db}{dt'}+\left\{\lfrac{1}{2}\frac{\dot{\xi}+\gamma_+
\dot{\bar{\xi}}}{\xi+\gamma_+\bar{\xi}}-\lfrac{i}{2}(\alpha-\gamma_+
\bar{\alpha})^2\frac{1}{\left(\xi+\gamma_+\bar{\xi}\right)^2}\right\}b=0.
\label{ss3.8}
\end{equation}
Integrating this equation, we find that 
\begin{equation}
b(t')={\cal N}\left(\xi+\gamma_+\bar{\xi}\right)^{-1/2}
\exp{\left[\lfrac{i}{2}(\alpha-\gamma_+\bar{\alpha})^2
\int\frac{dt'}{(\xi+\gamma_+\bar{\xi})^2}\right]}.\label{ss3.12}
\end{equation}
The integral in (\ref{ss3.12}) can be solved by noting that, if 
$\xi$ and $\bar{\xi}$ are solutions to Eq. (\ref{ode3.1}) then so too are 
$[\xi+\gamma_+\bar{\xi}]$ and  
$[\bar{\xi}+\bar{\gamma}_+\xi]$.  The latter two solutions satisfy the 
Wronskian
\begin{equation}
W(\xi+\gamma_+\bar{\xi},\bar{\xi}+\bar{\gamma}_+\xi)=-i(1-\bar{\gamma}_+
\gamma_+).\label{ss3.16}
\end{equation}

With these facts in mind, we obtain 
\begin{equation}
b(t')={\cal N}\left(\xi+\gamma_+\bar{\xi}\right)^{-1/2}\exp{\left[
-\lfrac{1}{2}\frac{(\alpha-\gamma_+\bar{\alpha})^2}{1-\bar{\gamma}_+
\gamma_+}\left(\frac{\bar{\xi}+\bar{\gamma}_+\xi}{\xi+\gamma_+\bar{\xi}}
\right)\right]},\label{ss3.20}
\end{equation}
and have the result
\begin{eqnarray}
\Psi_{\alpha,z}(x,t') & = & {\cal N}\left(\xi+\gamma_+\bar{\xi}\right)^{-1/2}
\exp{\left\{\frac{\alpha-\gamma_+\bar{\alpha}}{\xi+\gamma_+\bar{\xi}}
\left[x-\frac{\alpha-\gamma_+\bar{\alpha}}{2(1-\bar{\gamma}_+\gamma_+)}
\left(\bar{\xi}+\bar{\gamma}_+\xi\right)\right]\right\}}\nonumber\\
 &  & \hspace{1cm}\times\exp{\left[\lfrac{i}{2}\frac{\dot{\xi}+\gamma_+
\dot{\bar{\xi}}}{\xi+\gamma_+\bar{\xi}}\,x^2\right]}.\label{ss3.24}
\end{eqnarray}
The integration constant ${\cal N}$ will be fixed by normalization.

A more transparent expression for $\Psi_{\alpha,z}(x,t')$ can be obtained 
by first decomposing 
the exponential terms into their real and complex parts. 
After some algebra and normalizing,
\begin{eqnarray}
\Psi_{\alpha,z}(x,t') & = & 
\left(\frac{1-\bar{\gamma}_+\gamma_+}{2\pi {\bf Q}_3}
\right)^{\frac{1}{4}}\left(\frac{\bar{\xi}+\bar{\gamma}_+\xi}{\xi+\gamma_+
\bar{\xi}}\right)^{\frac{1}{4}}
\exp\left\{-\frac{1}{4}\frac
{(1-\bar{\gamma}_+\gamma_+)}
{{\bf Q}_3}\left[x-{\bf X}_3^+(\alpha,z)\right]^2\right\}
\nonumber\\
 &   & 
\times\exp
\left\{i\left[\frac{1}{4}\frac{{\bf R}_3}{{\bf Q}_3}x^2
+\frac{1}{2{\bf Q}_3}
\left(x-\lfrac{1}{2}{\bf X}_3^+(\alpha,z)\right){\bf X}_3^-(\alpha,z)
\right]\right\},
\label{ss3.28}
\end{eqnarray}
\begin{equation}
{\bf Q}_3 = (\xi+\gamma_+\bar{\xi})(\bar{\xi}+\bar{\gamma}_+\xi),~~~~
{\bf R}_3 = \dot{{\bf Q}}_3, \label{ssfns3.1}
\end{equation}
\begin{eqnarray}
{\bf X}_3^{+}(\alpha,z)  &=&  
  (\alpha-\gamma_+\bar{\alpha})(\bar{\xi}+\bar{\gamma}_+\xi)
  + (\bar{\alpha}-\bar{\gamma}_+\alpha)(\xi+\gamma_+\bar{\xi}), 
\nonumber \\
{\bf X}_3^{-}(\alpha,z)  &=&  -i\left[
  (\alpha-\gamma_+\bar{\alpha})(\bar{\xi}+\bar{\gamma}_+\xi)
  - (\bar{\alpha}-\bar{\gamma}_+\alpha)(\xi+\gamma_+\bar{\xi})\right].
\label{s88}
\end{eqnarray}
In Eq. (\ref{ss3.28}), the argument in the first exponential term is real, 
while the argument in the second is pure imaginary.  The wave function 
(\ref{ss3.28}) is clearly Gaussian. All the quantities defined in 
Eqs. (\ref{ssfns3.1}) and (\ref{s88}) are real.

To further simplify, first define the following quantities:
$$
\gamma_-=-{{\bar{z}}\over{|z|}}\tanh{|z|},~~~~~
\gamma_+= {{z}\over{|z|}}\tanh{|z|},~~~~~
\gamma_3=-\ln{\left(\cosh{|z|}\right)}, 
$$
$$
z=re^{i\theta}, ~~~~~ r=|z|, \nonumber
$$
\begin{equation}
Q_3=\frac{{\bf Q}_3}{(1-\bar{\gamma}_+\gamma_+)}=
\lfrac{1}{2}\left[\phi_3\cosh{2r} + \left(\phi_1e^{-i\theta} 
    +\phi_2e^{i\theta}\right) \sinh{2r}\right], 
\label{s89}
\end{equation}
\begin{equation}
R_3=\frac{{\bf R}_3}{(1-\bar{\gamma}_+\gamma_+)}=\dot{Q}_3,
\label{s90}
\end{equation}
\begin{equation}
X_3^{\pm}(\alpha,z)=
\frac{{\bf X}_3^{\pm}(\alpha,z)}{(1-\bar{\gamma}_+\gamma_+)},
\label{s91}
\end{equation}
where in Eq. (\ref{s89}) we have used Eq. (\ref{phi}). 

The resulting squeezed-state wave function is
\begin{eqnarray}
\Psi_{\alpha,z}(x,t') & = & \left(\frac{1}{2\pi Q_3}\right)^{\frac{1}{4}}
\left(\frac{\bar{\xi}+\xi e^{-i\theta}\tanh{r}}
     {\xi+\bar{\xi} e^{i\theta}\tanh{r}}\right)^{\frac{1}{4}}
\exp\left\{-\frac{1}{4Q_3}\left[x-X_3^+(\alpha,z)\right]^2\right\}
\nonumber\\
 &   & \times\exp\left\{
i\left[\frac{1}{4}\frac{R_3}{Q_3}x^2
           +\frac{1}{2Q_3}
\left(x-\lfrac{1}{2}X_3^+(\alpha,z)\right)
X_3^-(\alpha,z)\right]
\right\}.
\label{s92}
\end{eqnarray}

Combining Eqs. (\ref{s88}), (\ref{s89}), and (\ref{s91}), 
we can obtain the following relationships:  
\begin{equation}
X_3^{+}(\alpha,z)=X_3^{+}(\alpha)
\label{s93}
\end{equation}
\begin{equation}
X_3^{-}(\alpha,z)
=X_3^{-}(\alpha) \cosh{2r} +Y_3^-(\alpha,\theta)\sinh{2r},
\label{s94}
\end{equation}
where the $X_3^{\pm}(\alpha)$ are given by Eq. (\ref{csfns1.1}), and 
\begin{eqnarray}
Y_3^-(\alpha,\theta) & = & 
-i\left(\alpha\xi e^{-i\theta} - \bar{\alpha}\bar{\xi} e^{i\theta}\right) 
\nonumber \\
 & = & \left(\xi^o\xi e^{-i\theta}
                    +\bar{\xi}^o\bar{\xi}e^{i\theta}\right)p_o
      -\left(\dot{\xi}^o\xi e^{-i\theta}
                   +\dot{\bar{\xi}}^{\,o}\bar{\xi}e^{i\theta}\right)x_o.
\label{s95} 
\end{eqnarray}
To obtain the last equation, we have used Table 1.

When $z=0$, $2Q_3 = \phi_3$, $2R_3 = \dot{\phi}_3$, and
$X_3^{-}(\alpha,0)=X_3^{-}(\alpha)$.  Then  the $TO$ squeezed-state wave 
function, $\Psi_{\alpha,z}$, reduces to the coherent-state wave function, 
$\Psi_{\alpha}$, of Eq. (\ref{cs3.3}). 

In the Appendix we show, by way of example,  how for the ordinary 
harmonic oscillator this can be reduced to the ordinary 
squeezed-state wave function. 


\subsection{
$TM$ Squeezed-State Wave Functions}

To obtain the squeezed-state wave functions for $TM$ systems, 
we note that in general
\begin{equation}
\hat{\Theta}_{\alpha,z}(x,t)=(\Psi_{\alpha,z}(x,t')\circ t')(t).
\label{ss2.1}
\end{equation}
Hence, from Eq. (\ref{s92}), we have
\begin{eqnarray}
\hat{\Theta}_{\alpha,z}(x,t') 
& = & \left(\frac{1}{2\pi \hat{Q}_2}\right)^{\frac{1}{4}}
\left(\frac{\hat{\bar{\xi}}+\hat{\xi} e^{-i\theta}\tanh{r}}
     {\hat{\xi}+\hat{\bar{\xi}} e^{i\theta}\tanh{r}}\right)^{\frac{1}{4}}
\exp\left\{-\frac{1}{4\hat{Q}_2}
\left[x-\hat{X}_2^+(\alpha,z)\right]^2\right\}
\nonumber\\
 &   & \times\exp\left\{
i\left[\frac{1}{4}\frac{\hat{R}_2}{\hat{Q}_2}x^2
           +\frac{1}{2\hat{Q}_2}
\left(x-\lfrac{1}{2}\hat{X}_2^
+(\alpha,z)\right)\hat{X}_2^-(\alpha,z)\right]\right\},
\label{s97}
\end{eqnarray}
\begin{eqnarray}
\hat{Q}_2&=&
\lfrac{1}{2}\left[\hat{\phi}_3\cosh{2r} + \left(\hat{\phi}_1e^{-i\theta} 
    +\hat{\phi}_2e^{i\theta}\right) \sinh{2r}\right], 
\label{s98}  \\
\hat{R}_2&=&\lfrac{1}{2}\left[\hat{\dot{\phi}}_3\cosh{2r} 
        + \left(\hat{\dot{\phi}}_1e^{-i\theta} 
        +\hat{\dot{\phi}}_2e^{i\theta}\right) \sinh{2r}\right], 
\label{s99}
\end{eqnarray}
\begin{eqnarray}
\hat{X}_2^{+}(\alpha,z)&=&\hat{X}_2^{+}(\alpha),
\label{s100}  \\
\hat{X}_2^{-}(\alpha,z) &=&
\hat{X}_2^{-}(\alpha) \cosh{2r} +\hat{Y}_2^-(\alpha,\theta)\sinh{2r}.
\label{s101}
\end{eqnarray}
The quantities  $\hat{X}_2^{\pm}(\alpha)$ 
are  defined in  Eq. (\ref{csfns2.1}) and 
\begin{eqnarray}
\hat{Y}_2^-(\alpha,\theta) & = & 
   -i\left(\alpha \hat{\xi} e^{-i\theta}
                - \bar{\alpha} \hat{\bar{\xi}} e^{i\theta}\right)
\nonumber \\
 & = & \left(\hat{\xi}^o\hat{\xi} e^{-i\theta}
               +\hat{\bar{\xi}}^o\hat{\bar{\xi}}e^{i\theta}\right)p_o
      -\left(\hat{\dot{\xi}}^o\hat{\xi} e^{-i\theta}
        +\hat{\dot{\bar{\xi}}}^{\,o}\hat{\bar{\xi}}e^{i\theta}\right)x_o,
\label{s102} 
\end{eqnarray}
is a real quantity. 

When $z=0$, $2\hat{Q}_2=\hat{\phi}_3$, 
$2\hat{R}_2=\hat{\dot{\phi}}_3$, and 
$\hat{X}_2^{-}(\alpha,0)=\hat{X}_2^{-}(\alpha)$.  Then  the 
$TM$-squeezed-state wave function, $\hat{\Theta}_{\alpha,z}$, 
reduces to the coherent-state wave function, $\hat{\Theta}_{\alpha}$, 
of Eq. (\ref{cs2.1}).


\subsection{
$TQ$ Squeezed-State Wave Functions}

To obtain the squeezed state wave function for $TQ$ systems, we follow the 
same procedure that we employed for their coherent states.  It is 
straightforward to show that 
\begin{equation}
\Phi_{\alpha,z}(x,t)=e^{-i\nu D}\hat{\Theta}_{\alpha,z}(x,t)=e^{-\nu/2}
\hat{\Theta}_{\alpha,z}(xe^{-\nu},t).\label{ss1.1}
\end{equation}
Therefore, from Eq. (\ref{s97}) we obtain 
\begin{eqnarray}
\Phi_{\alpha,z}(x,t) 
& = & \left(\frac{1}{2\pi Q_1}\right)^{\frac{1}{4}}
\left(\frac{\bar{\Xi}_P+ \Xi_P e^{-i\theta} \tanh{r}}
{\Xi+ \bar{\Xi}_P e^{i\theta} \tanh{r}}\right)^{\frac{1}{4}}
\exp\left\{-\frac{1}{4}\frac{R_1}{Q_1}
\left[x-X_1^+(\alpha,z)\right]^2
\right\}\nonumber\\
 &   & \times\exp\left\{
i\left[\frac{1}{4}\frac{R_1}{Q_1}x^2
+\frac{1}{2Q_1}
\left(x-\lfrac{1}{2}X_1^+(\alpha,z)\right)X_1^-(\alpha,z)
\right]\right\},
\label{s103}
\end{eqnarray}
\begin{eqnarray}
Q_1 & = & (\Xi_P+\gamma_+\bar{\Xi}_P)(\bar{\Xi}_P+\bar{\gamma}_+\Xi_P),
\label{s104} \\
R_1 & = & (\Xi_P+\gamma_+\bar{\Xi}_P)(\bar{\Xi}_X+\bar{\gamma}_+\Xi_X)
+(\Xi_X+\gamma_+\bar{\Xi}_X)(\bar{\Xi}_P+\bar{\gamma}_+\Xi_P), 
\label{s105}  
\end{eqnarray}  
\begin{eqnarray}
X_1^{+}(\alpha,z) & = & X_1^{+}(\alpha), 
\label{s106}  \\
X_1^-(\alpha,z) & = & 
X_1^{-}(\alpha) \cosh{2r} +Y_1^-(\alpha,\theta)\sinh{2r}. 
\label{s107}
\end{eqnarray}
The quantities  $X_1^{\pm}(\alpha)$ are defined in  Eq. (\ref{csfns1.4}) 
and 
\begin{eqnarray}
{Y}_1^-(\alpha,z) & = & -i\left(\alpha {\Xi_P} e^{-i\theta}
                - \bar{\alpha} {\bar{\Xi}}_P e^{i\theta}\right)
\nonumber \\
 & = & \left( {\Xi}_P^o {\Xi}_P e^{-i\theta}
               +{\bar{\Xi}}_P^o {\bar{\Xi}}_P e^{i\theta}\right)p_o
      -\left( {\Xi}_X^o {\Xi}_P e^{-i\theta}
           + {\bar{\Xi}}_X^{\,o} {\bar{\Xi}}_P e^{i\theta} \right)x_o 
\label{s108} 
\end{eqnarray}
is a real quantity.

When $z=0$, $2Q_1 =C_{3,T}$, $2R_1=\hat{\dot{\phi}}_3$, and
$X_1^{-}(\alpha,0)=X_1^{-}(\alpha)$.  Then  the $TQ$ squeezed-state wave 
function, $\Phi_{\alpha,z}$, reduces to the coherent-state wave function, 
$\Phi_{\alpha}$, of Eq. (\ref{cs1.5}). 


\section{Summary}

In the above analysis, we have provided details about the construction of 
Lie symmetry operators for the oscillator algebras of the three classes of 
systems, represented by $TO$ Hamiltonians, $TM$ Hamiltonians, and 
$TQ$ Hamiltonians, given the constraint that the 
$TM$ Hamiltonians contain only $P^2$ and $X^2$ terms.  
Representation theory was then 
employed to construct the number states for the three systems.   
The number states are, in general, not eigenfunctions of the respective  
Hamiltonian; that is, they are not energy eigenstates.  This is a 
consequence of the fact that the Hamiltonians are not, 
in general,  an operator in the 
corresponding oscillator algebra.  Therefore, the extremal state in 
each set of number states is not necessarily a ground-state wave function.

Starting
from the extremal states and the ladder operators of the oscillator 
algebras for each of the $TO$, $TM$, and $TQ$ systems, 
the appropriate 
displacement and squeeze operators could 
then  be constructed to calculate 
their respective coherent states and squeezed states.
However, for the squeezed states it was simpler to 
solve  first-order partial differential equations. 

Our results generalize the discussions of the two specific 
$TM$ time-dependent Schr\"odinger equations given in  Refs. \cite{ex1,ex2}.
We have shown how to handle, in general, all systems of the type given
in Eq. (\ref{genH2}).  Thus, the method can be straight-forwardly applied
to other special cases \cite{mc}.  


\section*{Acknowledgements}

MMN acknowledges the support of the United States Department of 
Energy.  DRT acknowledges
a grant from the Natural Sciences and Engineering Research Council 
of Canada.


\section*{Appendix: Ordinary Squeezed States}

The squeezed-state wave functions 
for the harmonic oscillator {\cite{nt1,nt2}} 
can be calculated using Eq. (\ref{ss3.28}) and Table 1 with 
$\xi(t)=(2\omega)^{-1/2}e^{i\omega t}$.  Here, for 
convenience, we have chosen $t_o=0$ and dropped the prime on $t$. 
The wave functions are  
\begin{eqnarray}
\Psi_{\alpha,z}(x,t) & = & \left(\frac{\omega}{\pi}\right)^{1/4}
\left(\frac{1}{\cosh{2r}+\cos{[2\omega t-\theta}]\sinh{2r}}
\right)^{\frac{1}{4}}\nonumber\\
 &   & \times\left(\frac{e^{-i\omega t}\cosh{r}+e^{i[\omega t -\theta]}
\sinh{r}}{e^{i\omega t}\cosh{r}+e^{-i[\omega t -\theta]}
\sinh{r}}\right)^{\frac{1}{4}}\nonumber\\
 &   & \times\exp\left\{
-\frac{1}{2}\frac{\omega\left[x-
\left(\frac{p_o}{\omega}\sin{\omega t}+x_o\cos{\omega t}
\right)\right]^2}
{\cosh{2r}+\cos{[2\omega t -\theta]}\sinh{2r}}\right\}
              \nonumber\\
 &   & \times\exp\left\{-\frac{i}{2}\frac{\omega x^2\sin{[2\omega t -\theta]}
\sinh{2r}}{\cosh{2r}+\cos{[2\omega t -\theta]}\sinh{2r}}\right\}\nonumber\\
 &   & \times\exp\left\{i\frac{\omega
\left[x-\frac{1}{2}
\left(\frac{p_o}{\omega}\sin{\omega t}+x_o\cos{\omega t}
\right)\right]}
{\cosh{2r}+\cos{[2\omega t
         -\theta]} \sinh{2r}} 
\right.
\nonumber\\
 &   & \left.\hspace{1.5cm}\left[\frac{p_o}{\omega}
           \left(\frac{}{}\cos{\omega t}\cosh{2r}
  +\cos{[\omega t-\theta]}\sinh{2r}\right)
\right.\right.
\nonumber\\
 &  & \left.\left.\hspace{1.5cm}-x_o\left(\sin{\omega t}\cosh{2r}
   \frac{}{}
  -\sin{[\omega t-\theta]}\sinh{2r}\right)\right]\right\}. 
\label{hoss}
\end{eqnarray}
For $t=\theta=0$, and where $s=\exp{r}$,
\begin{equation}
\Psi_{\alpha,r}(x,0) = \left(\frac{\omega}{\pi s^2}\right)^{1/4}
\exp\left\{-\frac{1}{2}\frac{\omega(x-x_o)^2}{s^2}\right\}
\exp\left\{ip_ox\right\}\exp\left\{-\lfrac{i}{2}p_ox_o\right\}.
\end{equation}



\end{document}